\documentclass[a4paper]{article}

\usepackage{setspace} 

\usepackage[english]{babel}
\usepackage[round]{natbib}
\usepackage{alltt, macros,xspace,amsmath,amssymb,stmaryrd,wasysym,yfonts,listings,ifthen,
amssymb,mathpartir,graphicx,float,fancyvrb}

\usepackage[usenames,dvipsnames]{color}

\newcommand{\term}[1]{\text{\bfseries #1}}
\newcommand{\nterm}[1]{\text{\tt #1}}

\newboolean{showcomments} 
\setboolean{showcomments}{true} 
\newcounter{commentnumber} 
\definecolor{nicered}{rgb}{.647,.129,.149}
\newcommand{\nb}[2]{ 
   \addtocounter{commentnumber}{1} 
   \ifthenelse{\boolean{showcomments}} 
   {
   {\color{nicered}\textsf\small$\blacktriangleright$\textit{#1: #2}$\blacktriangleleft$}}{}}

\def \lz{\bar{0}}
\def \lo{\bar{1}}
\def \lt{\bar{2}}
\def \li{\bar{i}}

\def \ln{\bar{n}}

\begin{document}

\pagenumbering{arabic}

\title{ Distributed Quantum Programming }

\author{\begin{tabular}[t]{c@{\extracolsep{8em}}c} 
Ellie D'Hondt & Yves Vandriessche \\\\
       \multicolumn{2}{c}{Software Languages Lab} \\
       \multicolumn{2}{c}{Vrije Universiteit Brussel} \\\\
       \multicolumn{2}{c}{\small{10F719, Pleinlaan 2, 1050 Elsene, Belgium}}\\
       \multicolumn{2}{c}{\small{tel. +32 629 11 05, fax. +32 629  35 25}}\\
       \multicolumn{2}{c}{\small{\{Ellie.DHondt, Yves.Vandriessche\}@vub.ac.be}}\\
\end{tabular}}

\maketitle

\begin{abstract} In this paper we explore the structure and
  applicability of the Distributed Measurement Calculus (DMC), an
  assembly language for distributed meas\-ure\-ment\--based quantum
  computations. We describe the formal language's syntax and
  semantics, both operational and denotational, and state several
  properties that are crucial to the practical usability of our
  language, such as equivalence of our semantics, as well as
  compositionality and context-freeness of DMC programs. We show how
  to put these properties to use by constructing a composite program
  that implements distributed controlled operations, in the knowledge
  that the semantics of this program does not change under the various
  composition operations. Our formal model is the basis of a quantum
  virtual machine construction for distributed quantum computations,
  which we elaborate upon in the latter part of this work. This
  virtual machine embodies the formal semantics of DMC such that
  programming execution no longer needs to be analysed by hand. Far
  from a literal translation, it requires a substantial concretisation
  of the formal model at the level of data structures, naming
  conventions and abstraction mechanisms. At the same time we provide
  automatisation techniques for program specification where possible
  to obtain an expressive and user-friendly programming environment.
\end{abstract}

\subsection*{}
\textbf{Keywords: } quantum computing, measurement-based quantum
computing, distributed computing, formal models,
virtual machines.



\section{ Introduction }
During the last decennia, quantum information has managed to become a
significant field of research in the exact and applied sciences.
Although it is a relatively new discipline one can currently discern
several sub-disciplines such as quantum cryptography, information
theory, computability, error correction, fault tolerance, computations
and of course there is also the experimental research on the
construction of quantum computers~\citep{Nielsen00}. Nevertheless, the
development of quantum information as a proper computational domain of
computer science is lagging behind. Indeed there is no such thing as a
quantum computational paradigm. By this we mean a framework in which
quantum problems can be expressed and solved in terms of data
structures, algorithms, techniques such as abstraction, all of these
wrapped up in an associated programming language. Paradigm building
has proved to be an extremely useful approach in computer science. It
has led to theoretically equal but practically very different
programming frameworks, such as functional, imperative, logic and
object-oriented programming. For this reason we expect this approach
to be crucial also in developing quantum programming paradigms.

The first step in paradigm building is to construct a low-level
quantum programming language and determine its properties. By
low-level we mean that we need to define syntactical expressions for
each physically allowed quantum operation: preparation, unitary
transformation, measurement, combined with classical control
expressions if so desired. The syntax in its own is not the goal, but
rather a means by which to facilitate investigations with computer
science techniques. First, we need to determine the functionality of a
quantum program, its \emph{semantics}. The most obvious way to do this
is the operational semantics, the most practical is the denotational
semantics. While in the former a program's meaning is given as a
sequence of state-changing operations, in the latter it is instead a
mathematical object. Paramount is linking both, so that one can use
whichever in future analyses. Through a programming language's
semantics one can investigate notions such as composition and
context-freeness. These properties are crucial when one wants to build
more complex programs. Indeed, they allow the semantics of these
larger programs to be built up from that of smaller components using
rules for composition, rather than having to be determined from
scratch. While these properties may seem obvious, computer science is
littered with examples where they were mistakenly taken to be true,
leading to problems in programming language development (see for
example \citep{Brock81}).

Recent advances in quantum communication and cryptographic
computations motivate the need for a programming paradigm centred on
\emph{distributed} quantum computations. In a distributed system one
has concurrently acting agents in separated locations, operating
locally on a quantum state, which may be entangled over agents, and
coordinating their actions via communication. Formal frameworks for
distributed quantum computation have only very recently begun to
appear. Typically, these are a combination of classical process
theory, which formalises notions of concurrency and communication, the
quantum circuit model, i.e. local operations are unitary
transformations of an agent's qubits, and given initial shared
entanglement. First, there is the work in~\citep{Lalire04,Jorrand05},
which is directly built upon classical process calculi. While this
model profits from being closely related to existing classical models,
the disadvantage is that it is hard to focus on quantum behaviour. A
different approach is advanced in the model known as
\emph{communicating quantum processes} or CQP~\citep{Gay04,Gay05},
where the typical communication primitives of process calculi are
combined with computational primitives from QPL~\citep{Selinger03}. The
basic model of CQP is more transparent. This model serves as a basis
for the development of formal verification techniques, in particular
for proving the security of larger scale quantum communication
systems. In related work, a probabilistic model-checking tool built
upon an existing automated verification component is
developed~\citep{Gay05}. There is also the work in~\citep{Adao05}, which
is specifically tailored to security issues in cryptographic
protocols. In our work, however, we are much more interested in the
expressiveness of quantum distributed computations and the behavioural
properties of distributed primitives. In a way, we take the inverse
approach, assuming that computations are well-defined and
investigating what programming concepts are at work, instead of the
other way around.

While the fact that formal verification tools for distributed quantum
computation are currently under development may suggest that a mature
distributed paradigm already exists, this is actually not the case.
Distributed protocols are still very much conceived on intuition, and
there is no good notion or formalisation of the programming concepts
that are at work. We therefore require adequate formal tools with
which to explore and evolve the distributed quantum computing
paradigm. In this paper we explore the structure and applicability of
the Distributed Measurement Calculus (DMC) ~\citep{DHondt05,Danos05b},
an assembly language for distributed meas\-ure\-ment\--based
quantum computations. We describe its syntax and semantics, both
operational and denotational, and state several properties that are
crucial to the practical usability of our language, such as
equivalence of our semantics, as well as compositionality and
context-freeness of DMC programs. We show how to put these properties
to use by constructing a composite program that implements distributed
controlled operations, and demonstrate that the semantics of this
program does not change under the various composition operations. Finally, we elaborate on an implementation for the DMC language which we developed under the form of a virtual machine, a platform-independent programming environment
that abstracts away details of the underlying hardware or operating
system. This virtual framework is a crucial step in providing DMC as an experimentation platform for distributed quantum computations, as reasoning within the formal model by hand proves quite cumbersome for even small computations. 
At the basis of our work lies the measurement calculus~\citep{Danos04b}, a
low-level quantum programming language for measurement-based quantum
computations from which we explore the distributed paradigm. Next to
the obvious advantage of starting from a proper formal framework,
measurement-based models are well-suited as a starting point for
distributed quantum computations because they are inherently
distributed. What is important here is that the well-known physical
framework of quantum computation is ported to an equivalent computer
science framework, opening up the field towards investigations from
this branch of science as well.

The structure of this paper is as follows. In the next section we
discuss the syntax of our language, while Sec.~\ref{sec:semantics}
covers the semantics of DMC and lists its properties. Sec.~\ref
{sec:app} applies the previous to a concrete example, a composite
protocol implementing distributed controlled gates. We discuss the
implementation of the quantum virtual machine for DMC in
Sec.~\ref{sec:virtual}, and conclude in Sec.~\ref{sec:conclusion}. 
Some basic knowledge of quantum computation is assumed; for the reader
not familiar with the domain we refer to the
excellent~\citep{Nielsen00}. Our approach in this article is to explain
the notions of the model by example, rather than providing a series of
formal definitions which are hard to interpret and for which space it
too limited. However, we stress that the model is a rigorous one,
and, while this paper is a stand-alone document, refer the interested
reader to the appendix for the full formal semantics of the DMC language and ~\citep{DHondt05,Danos05b} for complete definitions.

\section{Syntax \label{sec:syntax}}

%
The language we propose is, broadly speaking, an assembly language for
distributed measurement-based quantum computations. It is an assembly
language in that it provides syntax only for the most basic operations
while at the same time being universal. It is measurement-based as we
build our language on top of the measurement calculus
(MC)~\citep{Danos04b}, an assembly language for measurement-based
quantum computations. The latter depart from the usual circuit-based
approach to quantum computing, where unitary transformations are the
driving force of computations. While measurement operations were long
seen as a necessary but disruptive part of quantum computing, in
algorithms such as teleportation they act as an essential part of the
computation. This gave rise to models where the computation is steered
by pre-established generic entanglement combined with measurements,
such as the one-way quantum computer \citep{Raussendorf02a}. Because
measurements are inherently probabilistic, correction operations are
required that are conditionally applied depending on previous
measurement outcomes, thus rendering the computation deterministic.
All of this is nicely captured in the measurement calculus. For this
reason, as well as the inherent distributive aspect of
measurement-based models, MC is an ideal basis from which to develop
our language.

We describe our language model in three layers. The base layer
consists of MC \emph{patterns}, which describe local quantum
computations. These are combined with distribution primitives into the
notion of \emph{agents} in the middle layer. Finally, agents and their
shared entanglement resources are bundled into \emph{networks} in the
top layer.

\paragraph{\tbf{Patterns.}}
In MC a pattern is defined by a sequence of commands together with
sets of qubits for working, input and output memory. As an example
consider the following pattern, which, as we explain below, implements
the Hadamard operation,

\EQ { 
\cH(1,2) := (\{1,2\},\{1\},\{2\},X_2^{s_1}\M X1\et12) \text{  .}
\label{hadamard}
}
All qubits are uniquely identified using numbers.  The first
argument  denotes that the pattern has a computation space of
two qubits, referenced by $1$ and $2$.  The next two arguments specify
the pattern's inputs and outputs. Working qubits that are not input qubits are initialised to $\p=\0+\1$.  The last argument is the pattern's command sequence, a list of operations taken from a basic set and executed from right to left, analogous to matrix applications. Subscripts denote qubit arguments  of the operation, while corrections are conditionally executed depending on their superscript. Concretely, one runs the pattern $\cH$ by preparing the first qubit in some input state $\gket$ and the second in state $\p$, then entangling both qubits with the controlled-$Z$ operation\footnote{A controlled-$Z$ operation on two qubits applies the $Z$ operation to the second qubit provided the first is set to 1.} to obtain $CZ_{12}(\gket \ox \p)$  (syntax: $\et12$). Next, the first qubit is measured in the $\ens{\p,\m}$ basis (syntax: $\M X1$), where $\m=\0-\1$. Finally, an $X$ correction is applied on the output qubit if the measurement outcome was $s_{1}$ (syntax: $\cx 2{s_1}$). Here $s_{1}$ defines a \textit{signal} -- simply 0 or 1 -- coming from the $X$-basis measurement on qubit 1 in $\M X1$. If  the signal is 0, the operation is not performed. Measurements are considered to be destructive, which is why qubit 1 does not appear in the output set. 

A general pattern is denoted $\cP(V,I,O,\cA)$, with computation space
$V$, inputs $I$ and outputs $O$ (together called the \emph{pattern
  type}), and command sequence $\cA$ that consists of
\emph{entanglements} $\et ij$, \emph{measurements} $\MS\al ist$, or
\emph{corrections} $\cx is$ or $\cz is$, where $i,j \in V$, $\al \in
[0,2\pi]$ and $s,t \in \mbb{Z}_{2}$. Allowed measurements are those in
the $XY$-plane of the Bloch sphere, and are specified by the angle
$\al$ ($\M X1$ is actually syntactic sugar for $\M 01$). Measurement
angles may also be conditioned by signals, written $\MS{\al}ist$, with
$(-1)^{s}\al+t\pi$ being the actual measurement angle. The four basic
instructions together with signal conditioning suffice to make the
model universal~\citep{Raussendorf02a,Danos04b}.

Patterns can be combined into larger ones to create arbitrary quantum computations. The \emph{sequential composition} of patterns $\cP_{1}=(V_{1},I_{1},O_{1},\cA_{1})$ and $\cP_{2}=(V_{2},I_{2},$ $O_{2},\cA_{2})$, with $O_{1}=I_{2}$, is defined as
\EQ{
\cP_{2}\cP_{1}:=(V_1\cup V_2, I_1, O_2,\cA_{2}\cA_{1}) \text{  ,}
\label{eq:composition}}
while the \emph{parallel composition} of the same patterns  is defined as
\EQ{
\cP_{1}\ox \cP_{2}:=(V_1\uplus V_2,I_1\uplus I_{2},O_{1}\uplus O_2,\cA_{1}\cA_{2}) \text{  .}
\label{eq:tensoring}}
Note that one can always rename qubits for parallel composition to become possible, while sequential composition also needs $I_{2}$ and $O_{1}$ to have the same cardinality.

As an example, here is the pattern to create a 3-fold GHZ-state $\ket{000}+\ket{111}$,
\EQ{
GHZ_{123}=(\ens{1,2,\hat{2},3,\hat{3}}, \cdot, \ens{1,2,3}, \cH(\hat{3},3)\et2{\hat{3}}\cH(\hat{2},2)\et1{\hat{2}})\;\text{,}
\label{ghz}
}
where $\hat{2}$ and $\hat{3}$ are working qubits.

MC is equipped with a powerful standardisation theorem which provides a procedure for bringing any pattern, such as the one above, into $EMC$-form, i.e. with entanglements first and corrections last. This is important from an experimental implementation perspective, and also corresponds nicely with the typical structure of a distributed quantum protocol where one starts out with a shared entanglement resource. In fact within MC we can already express distributed computations such as teleportation, 
\EQ { 
 (\{1,2,3\},\{1\},\{3\},X_{3}^{s_{2}}Z_{3}^{s_{1}}M_{2}^{X}M_{1}^{X}E_{12}E_{23})\text{  .}
  \label{tp}
}
While the  pattern describes the intended computation, we find no notion
of separate parties participating in the teleportation, and neither of the required communication between them. Purely by convention, and because everybody knows the protocol, we can say that qubits 1 and 2 belong to Alice and qubit 3
to Bob.  It is still hard to see if and what both parties have to communicate, e.g. Bob needs $s_1$ and $s_2$ but has no access to qubits 1 and 2, thus Alice needs to perform the measurements and
communicate the results to Bob.  We want this information to be explicit and obvious in the language, which in turn makes it easier to describe and reason about distributed quantum programs. To see that this is important, try to identify what is implemented with the following distributed protocol,
\EQ{
(\ens{0,\lz,1,\lo,2,\lt},\cdot,\ens{0,1,2}, \cz 2{s_{\lt}} \cz 1{s_{\lo}}\cz 0{s_{\lz}}\M{GHZ}{\lz\lo\lt}\et 0\lz \et 1\lo \et 2\lt)\text{  .}
\label{ESsimple}
} Here $\M{GHZ}{}$ is a pattern for GHZ-measurement. Note that pattern
(\ref{ghz}) implements the unitary transformation between the diagonal
basis and the GHZ-basis
$\ens{\ket{i}+\ket{\overline{i}}}_{i=0}^{2^{n}-1}$. Hence a
GHZ-measurement is executed by applying the inverse pattern followed
by a measurement in the diagonal basis. We will get back to this
pattern in Sec.~\ref{sec:app}.

\paragraph{\tbf{Agents.}}


An agent embodies a single computation node running in isolation in a
distributed algorithm, i.e. a processor or a party such as Alice or
Bob. Each agent has a local command sequence, which operates on a set
of resources contained within its environment. Agent instructions are
either measurement patterns or communication primitives. For example,
the teleportation pattern (\ref{tp}) really needs the following agent
definitions for Alice and Bob, respectively, \EQ{ \bA: \ens{1,2}.\cco
  s_2 . \cco s_1.M_{2}^{X}M_{1}^{X}\et12 \quad \text{and} \quad \bB:
  \ens{3}.X_{3}^{s_{2}}Z_{3}^{s_{1}}.\cci s_2 . \cci s_1 \text{ .} }
As we can see an agent definition consists of a \emph{type} in curly
braces, which specifies what qubits an agent owns, and an instruction
sequence. Next to local quantum operations, written in MC language, we
specify the exchange of the signals $s_{1}$ and $s_{2}$ between Alice
and Bob via the communication primitives $\cci$ and $\cco$. What is
missing here as compared to pattern (\ref{tp}) is the prior shared
entanglement $\et23$, as indeed it is impossible to factorise this
part of the protocol into agent definitions. There is no other option
than to specify shared entanglement separately in the network
definition for the full protocol, as we shall see below. Of course the
qubits are still local to the agents -- as reflected in the type --
only their description is not.

Formally, an agent is defined by an expression $\bA(i,o): Q.\cE$,
where the type $Q$ is a set of qubit references and $\cE$ is a finite
instruction sequence composed of pattern command sequences $\cA$,
classical message reception $\cci s$ and sending $\cco s $, where $s$
is a signal, and quantum reception $\qci q$ and sending $\qco q$,
where $q$ is a qubit reference. Corresponding communication actions
are synchronised, meaning that an agent executing a receive will pause
its program until the required agent has sent a message on the same
channel and vice versa. Classical inputs $i$ and outputs $o$ are used
in protocols such as superdense coding, and are not mentioned when
there are none. Also, working qubits required by patterns are
initialised to $\p$ as before and are not specified in the type $Q$.

\paragraph{\tbf{Networks.}}
A network of agents consists of several agents that execute their
command sequence concurrently. Typically, quantum resources are shared
between agents -- indeed most distributed quantum protocols benefit
from some type of shared entanglement being present prior to the start
of the protocol. We have no way of splitting the representation of
these states over agent definitions, and instead have to specify
agents separately in a network's definition. This is why a network
specification is more than just a collection of agent definitions, a
feature specific to quantum concurrent frameworks and absent in
classical concurrency. To keep track of how a shared resource is
distributed each agent's type $Q$ specifies which qubits of a shared
resource are local to that agent.

At this point we can finally write down the full teleportation
protocol, including all relevant distributed aspects. \EQ{\SP{
    TP :=\;& \;\;\bA: \ens{1,2}.\cco s_2 . \cco s_1.M_{2}^{X}M_{1}^{X}\et12  \\
   & | \;  \bB: \ens{3}.X_{3}^{s_{2}}Z_{3}^{s_{1}}.\cci s_2 . \cci s_1\\
   & \| \; \et23
}}
Analogous to process algebraic notation we use a vertical bar $|$ to
separate concurrently executing agents. Shared network resources are
specified after the double bar $\|$. For teleportation this is the
state $E_{23}$, which was produced and handed out to Alice and Bob
prior to the execution of the protocol. If so desired the
establishment of shared resources can itself be seen as a distributed
protocol involving a server agent polled for entanglement services. We
now likewise lift the pattern given in (\ref{ESsimple}) to a
distributed setting, obtaining the following
 \EQ{\SP{
ES:=\;&\;\;\bL  :\ens{\lz,\lo,\lt}.\mtt{c_{0}!}s_{\lz} .\mtt{c_{1}!}s_{\lo}.\mtt{c_{2}!}s_{\lt}.\M{GHZ}{\lz\lo\lt}\text{  .}\\
& |\; \bA_{0}  :\ens{0}.\cz 0{x_{\lz}}.\mtt{c_{0}?}x_{\lz}\\
& |\; \bA_{1}:\ens{1}.\cz 1{x_{\lo}}.\mtt{c_{1}?}x_{\lo}\\
&|\; \bA_{2}:\ens{2}.\cz 2{x_{\lt}}.\mtt{c_{2}?}x_{\lt}\\
&\| \; \et 0\lz \et 1\lo \et 2\lt\text{  ,}
\label{esn}
}}
where $\M{GHZ}{\lz\lo\lt}$ is a pattern executing a GHZ-measurement. This is the the entanglement swapping protocol for three agents \citep{Zukowski93,Bose98}, which produces a GHZ-state shared between $\bA_{0}$, $\bA_{1}$ and $\bA_{2}$. Because the resulting GHZ state is in the diagonal basis we have $Z$ rather than $X$ corrections in the network specification. 

Arbitrary networks are denoted
$\cN=|_{i}\bA_{i}(\bi_{i},\bo_{i}):Q_{i}.\cE_{i}\given \sig $. For
simplicity, we assume that networks of agents satisfy a number of
definiteness conditions which ensure that the computation is
well-defined. For example, an agent may only operate on qubits he
owns, and every communication event needs to have a corresponding dual
event in the network. Of course these issues are important but we are
glossing over them here as we assume checking these occurs in a
pre-compilation step rather than at runtime.
 
Our extensions to the measurement calculus have made distributed
notions explicit. We can now see directly from a protocol
specification what agents have to communicate to whom as well as which
assumptions are made about non-local entanglement resources. Also,
because each agent's instructions are expressed separately, we do not
impose any particular execution order for local quantum operations.
This makes it clear which parts of the protocols can and cannot run
concurrently.

\section{Semantics\label{sec:semantics}}




 In the previous section we established the syntax of DMC. The next
 step is to establish the meaning of a program written down in this
 language, i.e. its semantics. Formal semantics is a means of
 assigning objects to chunks of code so that one can reason with these
 objects rather than with the code itself. In this section we describe
 an operational as well as a denotational semantics for DMC programs
 and, with this in hand, state some important properties of our
 framework.

 \paragraph{\tbf{Definition.}} A network's operational semantics
 reflects how it affects the state of a distributed system on which
 the network is run. That is, we associate with each agent a local
 state (quantum and classical), and specify how the network
 specification updates these. Since we always have to take the shared
 entanglement resources into account the operational semantics does
 not decompose into state transformers for each of the agents
 separately. On top of this a crucial ingredient of the semantics
 specifies how quantum resources move throughout the system.
 Concurrency is not really an issue for the semantics because, due to
 the linearity of quantum mechanics, the order in which local
 operations are applied is unimportant. Hence we pick an order
 arbitrarily and derive the semantics for this case.

 Concretely, the operational semantics of a network is found by
 collapsing individual small-step semantical rules into one transition
 system. The full small-step semantics of DMC can be found in the appendix; it consists  of a
 number of quantum mechanical rules -- corresponding to measurement
 pattern commands -- and a number of communication rules. These affect
 the global quantum state ($EMC$ commands), the local resources
 (quantum communication), and the local memory ($M$ commands and
 classical communication). Since measurement is probabilistic, so are
 the small-step rules. The formal semantics of measurement pattern
 commands is well established~\citep{Danos04b,DHondt05}, and just needs
 to be lifted to the DMC setting. We elaborate here the semantics of
 typically distributed concepts introduced in the previous section. The formal semantics of MC relies on the component $\Ga$, the outcome
 map, which contains a number of bindings from signal names to
 measurement outcomes. In DMC the outcome map is lifted to a local
 agent memory recording also the values of classical messages.
 Concretely, a classical communication event between two agents has
 the following semantics, given that $\Ga_{b}$, the local memory of
 Bob, evaluates the name $y$ to the value $v$,
 \EQ{\SP{
&\;\bA:(Q_{a},\Ga_{a}) .\cE_{a}. \cci x \;|\; \bB:(Q_{b},\Ga_{b}).\cE_{b }.\cco y \\
\Lrar&\; \bA:(Q_{a},\Ga_{a}[x\rar v]).\cE_{a}\;|\; \bB:(Q_{b},\Ga_{b}).\cE_{b} \quad \text{  .}
\label{rendvcl}
}}
 Here $\Ga_{a}[x\dar v]$ means that we assign a new binding of $x$ to $v$ in the local memory of Alice. These rules are typically much harder to read and write down formally than to apply concretely. What we are saying here is that if a classical communication takes place, the value is looked up by the sending agent, and received and bound to a new name by the receiving agent, after which the computation ($\cE_{a }$ and $\cE_{b }$) continues.  Sending and receiving qubits over a quantum channel changes the types $Q$ of the agents involved.  An agent sending a qubit can no longer perform any operations on it, therefore the corresponding qubit reference is removed from $Q$, while it is added to the receiving agent. Formally, the rule for the exchange of a qubit with reference $i$ is given by
\EQ{\SP{
&\;\bA:(Q_{a},\Ga_{a}) .\cE_{a}. \qci i \;|\; \bB:(Q_{b},\Ga_{b}).\cE_{b }.\qco i \\
\Lrar&\; \bA:(Q_{a} \cup \ens{i},\Ga_{a}).\cE_{a}\;|\; \bB:(Q_{b}\bs\ens{i},\Ga_{b}).\cE_{b} \;\text{  .}
\label{rendvq}
}} It goes without saying that the way in which the actual
communication of a qubit is implemented is more intricate than this
simple rule. However, by providing quantum communication as primitive
syntax we are precisely choosing not to get into these implementation
matters. Both rules mentioned here do not affect the global quantum
state, which is why it is not mentioned. This is not the case for the
small-step semantics for pattern commands, which moreover may be
probabilistic. The full small-step semantics can be found in the appendix or
in~\citep{DHondt05,Danos05b}.

We obtain the operational semantics by defining computation paths in
the usual way and gathering all small steps in a computation path in
one big step from initial to final state in that path. As such the
operational semantics of a quantum distributed network is essentially
a probabilistic transition system (PTS). However, since resource
allocation is crucial, we have to augment this PTS with information on
how qubits move throughout the network. This is formalised as a type
signature. We denote the operational semantics of a network $\cN$ by
$\ops{\cN}$. For example, the operational semantics of $TP$ is given
by the deterministic transition system \EQ{ \ops{TP}:
  (\ens{1},\cdot)\rar(\cdot,\ens{3})\; . \; \rho_{1}\Lrar
  \rho_{3}\text{ ,}
 } 
 where $\rho$ is the density matrix specifying the input quantum state
 to be teleported and subscripts indicate qubit systems. The
 operational semantics of the $ES$ network given in (\ref{esn}) is
 also deterministic.
 \EQ{ 
 \ops {ES}: (\cdot,\cdot,\cdot,\cdot)\rar(\cdot, \ens{0} ,\ens{1} ,\ens{2})\;.\; \mbf{0} \Lrar GHZ_{012}^{D} \text{ ,}
 \label{esnsem}
} where the superscript $D$ denotes that the resulting state is in the
diagonal basis. Note that we only write real quantum inputs in the
type while not mentioning entanglement resources. The good thing about
operational semantics is that in principle it can be derived
automatically by induction on the small-step rules.

Denotational semantics is a second means of assigning a formal meaning
to a chunk of code, this time by way of mathematical objects. If we
look at the skeleton of a distributed protocol, i.e. the equivalent
non-distributed pattern, we find a multi-local probabilistic quantum
operation, which is mathematically represented by (a special type of)
completely positive map (CPM). Since protocols with different
distribution of resources are observationally different, we have to
pair this CPM with a function mapping input to output resources,
formally represented by a type signature. We denote this type of
semantics by $\des{\cN}$ for a network $\cN$. For example, the
denotational semantics of $TP$ is given by the map \EQ{ \des{TP}:
  (\ens{1},\cdot)\rar(\cdot,\ens{3})\;.\; \cI\text{ ,}
}i.e. $TP$ implements the identity map from qubit 1 to qubit 3. Here
we see the importance of the type signature in specifying the
semantics. The subtle difference between both types of semantics
becomes more apparent once one starts investigating more complex
protocols involving mixed states and probabilistic semantics.

As we see from the examples both types of semantics are quite similar,
and indeed we have the following result, which we state without proof.
 \begin{proposition}
 There is a \emph{precise correspondence} between the operational and the denotational semantics of networks of agents, that is to say
\EQ{
\forall \cN_{1},\cN_{2}: \cN_{1}\equiv_{op}\cN_{2} \iff \cN_{1}\equiv_{de}\cN_{2} \text{  .}
}
\label{equivsem}
\end{proposition}
Two networks are semantically equivalent if they have the same
semantics.\footnote{For example, one can prove that qubit
  communication between two agents is semantically equivalent to the
  teleportation network.} While the equivalence between semantics may
seem obvious (certainly from the example), it is crucial to prove this
statement formally, as it allows one to switch between semantics where
appropriate without giving the issue further thought. We note that in
general this equivalence is not guaranteed.

\paragraph{\tbf{Building larger protocols.}} 
With syntax and semantics in hand we can now consider constructing
larger networks from smaller components. We consider parallel as well
as sequential composition of networks, denoted by $\ox $ and $\circ$
respectively. These operations are defined in the obvious way so we
will not spell out concrete definitions here. Suffice to say that one
needs to pay attention to agent as well as qubit names to ensure that
networks are connected as desired. More importantly we need to make
sure that these forms of composition are consistent with the
semantics, as is indeed the case.
 \begin{proposition}
The semantics of networks is \emph{compositional}, i.e.
\AR{
\sem{\cN_{2}\circ\cN_{1}}&=\sem{\cN_{2}} \circ \sem{\cN_{1}}\text{  .}\\
\sem{\cN_{1}\ox\cN_{2}}&=\sem{\cN_{2}}\ox\sem{\cN_{1}}
}
\label{compprop}
 \end{proposition}
 Here we have a first application of Prop.~\ref{equivsem}, as first we
 do not need to specify which type of semantics we mean in the
 statement of Prop.~\ref{compprop} , and second we can choose the most
 convenient semantics for the proof, which is the denotational
 semantics. While the compositions on the left hand side are at the
 level of agent and network definitions, the composition of network
 semantics on the right hand side is a functional one at the level of
 type signatures, PTS's and CPMs. For the full proof we refer
 to~\citep{DHondt05}; an example is given in Sec.~\ref{sec:app}.

 A second important notion when constructing larger protocols, closely
 related to compositionality, is that of \emph{contexts}. Informally
 speaking a context is a program with a ``hole'' in which a network
 specification can be inserted, typically denoted $C[\cdot]$. On some
 occasions, one finds that programs that are considered equivalent in
 isolation no longer behave in the same way when placed within the
 context of a larger program. This is particularly the case in
 concurrent systems. A historical example is the \emph{Brock-Ackerman
   anomaly}~\citep{Brock81}, by which it was realised that simple
 input-output behaviour is not enough to reason about equivalence of
 concurrent systems. In our framework contexts $C[\cN]$ are given by
 arbitrary network compositions $C$ in which our network $\cN$ is
 placed. What is different with ordinary compositionality is that the
 quantum resources of the network $\cN$ are no longer considered to be
 independent of those of its context. Indeed in Prop.~\ref{compprop}
 quantum resources are considered to be provided independently, such
 that composite networks operate on disentangled inputs. While this is
 sensible when each of the networks operate in isolation, it is less
 so when a network is only one factor in a complex compositional
 structure. Hence we first need to show that our semantics is
 independent of so-called \emph{entanglement contexts}. This is a
 consequence of the following proposition, which we state here without
 proof. \begin{proposition} Suppose
   $\cL:\rho_{A}\lrar\rho_{B}=\sum_{k}L_{k}\rho_{A}L_{k}^{\dag}$ is a
   completely positive map. Then for all quantum states $\rho_{AC}$
   applying $\cL$ to the $A$-part of $\rho_{AC}$ results in \EQ{
     \rho_{BC}=\sum_{k}(L_{k}\ox I_{C})\rho_{AC}(L_{k}^{\dag}\ox
     I_{C}) . \label{extcpmap} } \label{entcont} \end{proposition} The
 proof, though easy, is not trivial. Using this proposition and
 compositionality we have the following important
 result.\begin{corollary} Equivalence holds in arbitrary contexts,
   that is, if $\cN_{1}\equiv \cN_{2}$, then for all network contexts
   $C[\cdot]$ we have that $C[\cN_{1}]\equiv C[\cN_{2}]$. \label{cont}
 \end{corollary}
  
\section{Applications\label{sec:app}}
We now show how all of the formal ingredients can be put to use in a
concrete example. The distributed primitive we will investigate is
that of a \emph{distributed remotely controlled
  gate}~\citep{Yimsiriwattana04}. Near-future quantum computers are
expected to have only a limited number of qubits per machine. Even in
quantum simulating environments the current qubit limit is only about
36~\citep{DeRaedt06}. Hence one can imagine that quantum computations
need to proceed much like cluster computations today, with resources
spread over different processors in order to make them feasible. One
common situation would be where the central processor needs to execute
a controlled operation with the target qubits spread over a group of
agents. Following the ideas in \citep{Yimsiriwattana04}, once we have a
GHZ resource we can execute a distributed controlled gate. Calling the
central processor \bL (for Leader), and assuming there are two
subordinate processors \bA and \bB , the trick is to establish a
shared control qubit between target agents, which is achieved through
the \emph{share control} (SC) protocol as follows, \EQ{\SP{
    SC := \;&\;\;\bL:\ens{c,0}.\mtt{c_{1,2}!}s_{0}.\;\M X0 \et 0c \text{  .}\\
    &|\;\bA_{1}:\ens{1}.\cx 1{x_{0}}.\mtt{c_{1}?}x_{0}\\
    &|\;\bA_{2}:\ens{2}.\cx 2{x_{0}}.\mtt{c_{2}?}x_{0}\\
    &\|\; GHZ_{012}^{D}
\label{asn}
}} Here qubit $c$ is the input control qubit which is in the state
$\al\0+\ba\1$. That the protocol indeed establishes its goal may be
seen from its semantics, which can be derived unambiguously to be
 \EQ{ \SP{
 \ops {SC}&: (\ens{c} ,\cdot, \cdot)\rar(\ens{c} ,\ens{1} ,\ens{2})\\
&\;(\al\0+\ba\1)_{c} \Lrar (\al\ket{000}+\ba\ket{111})_{c12} \text{ .}
 \label{asnsem}
}} Once this type of shared entanglement is in place each target agent
just has to execute a local controlled unitary gate with as control
its qubit in the shared entanglement resource. Note that because we
have context-independence of the semantics, target qubits may be
entangled over different agents. However, in order for the
distribution approach to be possible at all, the controlled unitary
$CU$ must have $U=U_{1}\ox U_{2}$ where, $U_{1}$ and $U_{2}$ operate
on a number of target qubits smaller or equal than the maximum
available qubits for each agent.

We see that the $SC$ protocol requires GHZ entanglement. It is
probably realistic to assume that in a quantum network each agent can
ask for Bell-state entanglement with the central server. It is not so
realistic to assume that groups of agents can demand direct GHZ
entanglement whenever they need it; rather, we expect GHZ entanglement
to be produced via the entanglement swapping protocol. Note that,
since we need GHZ entanglement between $\bL$, $\bA_{1}$ and $\bA_{2}$,
we need to compose $\bL$ with $\bA_{0}$ in the $ES$ protocol as
presented earlier in (\ref{esn}), that is \EQ{ \bL := \bA_{0} \circ
  \bL : \ens{0,\lz,\lo,\lt}. \cz
  0{x_{\lz}}.\mtt{c_{1}!}s_{\lo}.\mtt{c_{2}!}s_{\lt}.\M{GHZ}{\lz\lo\lt}\text{
    .} } What we are actually doing here is composing the $SC$
protocol with no shared resource with the $ES$ protocol to establish
the resource, and our semantics ensures that this is something we can
do unambiguously. Indeed, we have \EQ{\SP{
    &\sem{SC \circ ES}=\sem{SC}\circ \sem{ES} \\
    & = (\ens{c} ,\cdot,\cdot)\rar(\ens{c} ,\ens{1} ,\ens{2})\\
    &\qquad.\; (\al\0+\ba\1)_{c} \Lrar
    (\al\ket{000}+\ba\ket{111})_{c12} \text{,} }} with $\sem{SC}$ and
$\sem{ES}$ given in (\ref{asnsem}) and (\ref{esnsem}) respectively.

\paragraph{\textbf{More agents.}} The networks that we defined for implementing the distributed remote gate protocol can be generalised to $n$ agents. This requires generalised procedures for GHZ-measurement,  entanglement swapping and establishing a shared control qubit. First, an $n$-fold  GHZ state is produced by  generalising the 3-GHZ-pattern given in (\ref{ghz}). That is, through the pattern with no input qubits, output qubits $\ens{1,2,\dots,n}$ and an event sequence interleaving $E$ and $H$ operations, as follows,
\EQ{
GHZ_{1\dots n}=\cH(\hat{n},n)\et{(n-1)}{\hat{n}}\dots \et{3}{\hat{4}}\cH(\hat{3},3)\et2{\hat{3}}\cH(\hat{2
},2)\et1{\hat{2}} \;\text{,}
}
where the hatted qubits are again working qubits. Again, a GHZ-measurement is executed by applying the inverse pattern followed by a diagonal-basis measurement. This leads to the pattern
\EQ{
\M{GHZ}{1\dots n}=\M0{\hat{1}}\M0{\hat{2}}\dots\M0{\hat{n}}\et1{\hat{2}} \cH(2,\hat{2})\et{2}{\hat{3}}\cH(3,\hat{3})\et{3}{\hat{4}} \dots \et{(n-1)}{\hat{n}}\cH(n,\hat{n}) \;\text{,}
}
with input qubits the qubits $\ens{1,2,\dots,n}$ to be measured and with no outcome qubits. Using this sub-pattern we can establish GHZ-entanglement between the leader \bL and agents $\bA_{1}$ through to $\bA_{n}$ by the generalised entanglement swapping protocol \citep{Zukowski93,Bose98}, which has the following network specification,
\EQ{\SP{
ES := \; &\;\;\bL:\ens{0, \lz,\dots,\ln}.\cz 0{x_{\lz}}. (\mtt{c_{i}!}s_{\li})_{i=1}^{n}.\M{GHZ}{\lz\dots\ln}\\
&\;|_{i=1}^{n}\;\bA_{i}:\ens{i}.\cz i{x_{\li}}.\mtt{c_{i}?}x_{\li}\\
&\given\;  \ox_{i=0}^{n}\et i{\li} \text{  .}
\label{esnn}
}} This is just the generalisation of network (\ref{esn}) where we
have merged agents $\bL$ and $\bA_{0}$ for the purpose of establishing
a shared control qubit as before. The signal $s=s_{\lz}\dots s_{\ln}$
corresponds to a projection on the GHZ-state
$\ket{s}+\ket{\overline{s}}$. This network has the following
semantics,
 \EQ{ 
 \ops {ES}: (\cdot,\dots, \cdot)\rar(\ens{0},\ens{1},\dots,\ens{n})\;.\; \mbf{0} \Lrar GHZ_{0\dots n}^{D} \text{ .}
}
After establishing GHZ-entanglement between agents a shared control qubit is obtained through the following protocol.
\EQ{\SP{
SC := \;&\;\;\bL:\ens{c,0}.\mtt{c_{i}!}s_{0}.\:\M X0 \et 0c \text{  .}\\
&\;|_{i=1}^{n}\;\bA_{i}:\ens{i}.\cx i{x_{0}}.\mtt{c_{i}?}x_{0}\\
&\given\;  GHZ_{0\dots n}^{D}
\label{asnn}
}}
The semantics of this network is given by
 \EQ{ \SP{
 \ops {SC}&: (\ens{c} ,\cdot ,\dots, \cdot)\rar(\ens{c} ,\ens{1} ,\dots,\ens{n})\\
&\;(\al\0+\ba\1)_{c} \Lrar (\al\0^{\ox (n+1)}+\ba\1^{\ox(n+1)})_{c1\dots n} \text{ .}
 \label{asnnsem}
}} Control qubit $c$ can now indirectly control unitary operations at
the sites of all agents by having agent $\bA_{i}$ execute a local
pattern for a controlled unitary gate where the control is its qubit
$i$ and the targets are locally available qubits. Again, for this
approach to be possible the controlled unitary $CU$ must have
$U=U_{1}\ox\dots\ox U_{n}$ where each of the sub-unitary $U_{i}$ is
executable by agent $\bA_{i}$.

\section{Virtualisation \label{sec:virtual}}

In the above we introduced a formal language for distributed quantum
computation. Providing a number of tools for constructing higher-level
programs, it should be seen as a first step in a bottom-up
construction of a distributed quantum programming paradigm. However,
the previous section clearly shows how cumbersome it is to describe
and evaluate computations purely within the formal model. Rather,
these developments are only really useful when one thinks of the
language DMC in terms of a quantum virtual machine (QVM). A
\emph{virtual machine} is a platform-independent programming environment
that abstracts away details of the underlying hardware or operating
system. In our setting it is a low-level language abstraction layer
which executes DMC programs independent of the actual implementation
of quantum operations, which could be executed by any of several
existing quantum simulators or even by a physical quantum computer.
As a mediator between a set of low-level basic quantum gates and the
construction of more complex quantum programs, a QVM forms a crucial
layer in a tiered quantum computation architecture~\citep{Svore06}. It
is especially valuable if one is concerned with developing
higher-level distributed quantum computation languages through
experimentation and abstraction, for which a user-friendly and
flexible programming environment is invaluable.

Translating the formal model into a virtual machine has the obvious
benefits of automating program execution and composition. Furthermore,
low-level inspection during a pre-compilation step can automatically
determine a number of issues such as well-definedness of code. While
the formal language we have discussed in the above provides the
backbone for our virtual machine, its actual implementation is far
from trivial. Issues such as well-definedness of programs, naming of
variables within local computations as well as in larger program
contexts and efficiency all come into play in a more concrete sense
that is absent in the more abstract formal model. In what follows
below we give details on our QVM implementation. We first discuss the
QVM for sequential, MC-based computations, then extending the platform
further towards a distributed version thereof. We use the adjectives
\emph{formal} and \emph{virtual} to differentiate between similar
objects in the formal model and the virtual one whenever the context is
unclear.

\subsection{The quantum virtual machine\label{sec:exec-mc}}

The first step in the development of an execution environment for the
measurement calculus is translating formal measurement patterns into
structured data for the computer to work with. We represent this data
with \emph{symbolic} or \emph{s-expressions}~\citep{McCarthy:1960p228},
as popularised by the programming language Lisp.  At the same time we
chose to adhere insofar as possible to the notation used in the formal
model. Before discussing the full virtual syntax below, we give the
example of the Hadamard pattern from Eq.\eqref{hadamard}, which is
expressed virtually as
\begin{equation}\label{eq:pattern-data}
\cH:=
 \verb,((?i ?o) (?i) (?o) ((E ?o ?i) (M ?o 0) (X ?o (s ?i)))),
 \text{  .}
\end{equation}
Similarly to the formal setting we see a pattern expressed as a list
of three qubit sets ($V$, $I$, $O$) and a command sequence. Question
marks in qubit names indicate that they are variables, subject to
renaming and instantiation to concrete qubit references. Important to
note is that in the command sequence \emph{operations are executed
  from left to right}, in the formal model notation this was the
reverse. This change is due to both implementation reasons and
computer science tradition.

Our current execution environment is split in two separate layers: an
execution and composition layer. The execution layer takes a command
sequence as an input and is in charge of performing the quantum
operations it specifies. This command sequence is obtained after the
data representation of a measurement pattern is assembled into the
low-level language that the execution layer can understand --
essentially by replacing qubit variables with concrete qubit
references. Hence in terms of abstractness this layer lies below the
one in which ordinary patterns are defined. The composition layer, on
the other hand, adds an abstraction layer on top of pattern definition
in terms of the pattern composition structures from MC. It compiles
the composition of patterns into a single pattern data representation,
by merging arbitrary compositions of multiple patterns into a single
new pattern. In the rest of this section we discuss implementation
details of each of these components, focusing on respectively the
execution layer and its pattern assembler, as well as the composition
layer and its compiler.

\subsubsection{Execution layer\label{sec:qvm}}

One arrives at the execution layer whenever a pattern has to run,
either directly as a result of its definition being called or
indirectly after a composition of patterns was compiled to an
expression ready for execution. The execution layer consists of an
interpreter which directly executes each operation in its input
language. This interpreter is the core of our quantum virtual machine
(QVM), so called to stress the low-level and machine-generated nature
of its input language. It is essentially an automatic version of the
operational semantics described earlier in Sec.~\ref{sec:semantics}
and specified in full in the appendix.  The input language syntax for
the QVM is obtained by turning MC notation into an s-expression form,
specified as follows in a BNF notation:
\begin{equation} \label{eq:mc-bnf}
\begin{tabular}{l l l}
  \nterm{<sequence>}     & ::= & \term{(} \nterm{\{ <instruction> \}} \term{)}\\
  \nterm{<instruction>}  & ::= &\nterm{<correction> | <measurement> | <entanglement>}\\
  \nterm{<correction>}   & ::= &\term{( X}\nterm{ <quref> [ <signal> ]} \term{)} \nterm{|}\\
                                           &  &\term{( Y}\nterm{ <quref> [ <signal> ]} \term{)}\\
  \nterm{<measurement>}  & ::= &\term{( M}\nterm{ <quref> <angle> [ <signal> ]
    [ <signal>  ]} \term{)}\\
  \nterm{<entanglement>} & ::= &\term{( E}\nterm{ <quref> <quref>} \term{)}\\
  \nterm{<signal>}       & ::= & \term{0} \nterm{|} \term{1} \nterm{| <input-name> |}\\
  &  &\term{( s}\nterm{ <quref>} \term{)} \nterm{|} \term{(} \nterm{+ <signal> \{ signal \}} \term{)}\\
\end{tabular}
\end{equation}
where the symbols in boldface indicate syntax (round brackets and literals), while square and curly brackets are meta-syntax symbols
which denote option (zero or one occurrence) and repetition (occur
any finite number of times) respectively.
The only real difference with the formal syntax is the notation
{\term{( s}\nterm{ <quref>} \term{)} for signals, referring explicitly
  to measurement outcomes rather than relying on the naming convention
  $s_n$ to denote the outcome on qubit $n$. The execution layer
  interpreter takes any command sequence written in this language and
  executes it according to the semantics of MC. Note that in this way
  we already obtain a low-level quantum programming language. For
  example we can write and execute the Hadamard gate with the
  following command sequence:

\begin{equation}\label{eq:asm}
\verb,((E 0 1) (M 0 0) (X 1 (s 0))),
\text{  .}
\end{equation}
Transforming a pattern definition Eq.~\eqref{eq:pattern-data} to its
assembled version Eq.~\eqref{eq:asm} is relatively simple.
Concretely, the pattern assembler replaces each qubit variable name in
a pattern's command sequence by concrete qubit references.  For every
distinct variable name used we choose a unique integer and replace
every occurrence of the variable by it. At this stage only the input
qubit set is used, in particular for initialisation of the (non-input)
auxiliary qubits to the default $\p$-state. As we shall see below both
input and output set qubit sets are used when combining patterns in
the composition layer.

Practically, the interpreter takes besides a command sequence also an
\emph{environment}, which contains the current state of
execution. After an initialisation step the first operation in the
sequence is evaluated, after which the interpreter recursively calls
itself with an environment modified by this operation.  Just as in the
formal model, the environment for the interpreter consists of a
quantum as well as a classical part, respectively denoted $Q$ and
$\Gamma$ below. They are  defined as follows, where $T$ stands for \emph{tangle}.

\begin{align} 
\label{eq:qstate}
  \Gamma ::= & \; (o,i)  \\
 Q::= & \; T \, |\,  T \otimes Q \\
 T::= & \; (\{q_1,q_2,\ldots q_n\},\ket{\psi_{q_1 q_2 \ldots q_n}}), 
n \in \mathbb{N}
\end{align}
Differences with the formal model arise from practical
considerations. The classical state $\Gamma$ is split into the outcome
map $o$ and the input map $i$ to simplify lookup of qubit and input
names respectively.  Effectively, evaluating an expression of the form
\lstinline$(s n)$, where $n$ is a number, becomes a lookup in $o$.
  In the formal model the
quantum state (called $\rho$ in the semantics) is presented as a
single qubit state by creating a tensor product of all computation
qubits during initialisation. Given the computational explosion of
operations on larger tensors, this is highly impractical in a
classical simulation environment. For this reason we introduce the
concept of a tangle, a set of qubits which is known to be disentangled
from the full quantum state and thus allowing a more compact
representation. Hence in the QVM the full qubit state $Q$ is composed
of a number of individual tangle objects $T$, which are, evidently,
updated during execution.  
Concretely, during initialisation a new tangle
$T_i:(\{q_i\}, \gket)$ is created for each qubit, where $q_i$ a single
qubit reference, and auxiliary qubits have $\gket=\p$. After
interaction due to an entanglement operation $E_{q_i q_j}$ qubit
references $q_i$ and $q_j$ are put into the same tangle $T_{ij\dots}:
(\{q_i,q_j,\dots\},\ket{\psi_{q_i q_j \dots}})$, while the original
tangles $T_i$ and $T_j$ are destroyed. Here the dots represent other
qubits which have interacted with qubits $i$ or $j$ earlier in the
computation.  After initialisation each command in the command
sequence is executed closely following the semantics of the formal
model, albeit updated to the slight variations in auxiliary structures
as exhibited above. In our proof-of-concept implementation, we have
implemented the semantics directly using numerical linear
algebra. This implementation was based on an Lisp-based simulation
environment developed earlier in~\citep{Desmet06}.

\subsubsection{Composition layer\label{sec:compilation}}

The language understood by our execution layer is essentially MC
without composition operators. Writing larger quantum algorithms like
this by hand quickly becomes a tedious and error-prone process. The
measurement calculus introduces measurement patterns and pattern
composition which simplify this task. Pattern definition enables local
reasoning while composition allows the creation of more complex
programs out of smaller parts.  In this way more general patterns can
be defined by using parameters, and algorithms can be programmed by
composing them out of smaller building blocks. All these abstractions
are taken care of by the composition layer which we explain now. Its
compiler effectively machine-generates command sequences in the
language we see above.

Patterns are command sequences where all qubit names are categorised
as either input, output or working qubits.  Defining patterns is a
straightforward process: the command sequence syntax is the same as
the execution layer language, but instead of concrete qubit references
variable names may be used. An example is shown in
Eq.~\eqref{eq:pattern-data}.  On top of this a pattern definition can
be parametrised, which is necessary for patterns such as the
$\cJ(\alpha)$-pattern~\citep{Danos04b}.

By using qubit variables we facilitate the rewriting of concrete qubit
names, which in turn allows patterns to be composed by concatenating
their command sequences while sharing the correct qubit names.
Composing two patterns into a larger one is defined in
Eqs.~\eqref{eq:composition} and \eqref{eq:tensoring}. These definitions rely
on the programmer to prepare the composition by renaming qubit names
and, in case of sequential composition, tensoring identity patterns
where needed to make sure the condition $V_1 \cap V_2 = O_1 = I_2$ is
met. In a setting where we wish the programmer to be able to compose
arbitrary patterns in intricate configurations it is crucial to
automatise this renaming process. Our pattern compiler does precisely
that. It allows the programmer to declare his intentions in
non-trivial composition cases, whereas a standard rule is applied in
absence thereof. We explain the general composition case first, as we
define the standard rule in terms thereof later on.

As an example let us consider the controlled-not pattern $\ctR\cX$,
which is defined in terms of the $\cH$ and $\ctR
\cZ=(\ens{1,2},\ens{1,2},\ens{1,2},\et12)$ patterns, as follows.

\begin{equation} \label{eq:original-cx}
\ctR \cX := (\cI(1) \ox \cH(3,4)) \circ \ctR\cZ(1,3) \circ  (\cI(1) \ox \cH(2,3))
\end{equation} 
where $\cI$ is the identity pattern, used as filler to match the
number of in and output qubits. This composition pattern simply
follows from the analogous matrix identity. The same intent can also
be expressed without using concrete qubit names as follows

\begin{equation} \label{eq:new-cx}
\SP{
\ctR \cX :=\{&(\cI(q_8) \ox \cH(q_6,q_7)) \circ \ctR\cZ(q_5,q_4) \circ (\cI(q_3) \ox \cH(q_1,q_2))\\
& \text{with\;}  q_2=q_4, q_3 = q_5, q_4 = q_6, q_5 = q_1)\}
\text{,}}
\end{equation}
which is just the textual version of Figure~\ref{fig:composition} (in
fact, as we shall see below, our framework also provides a graphical
tool for specifying constraints, much as in the figure). 
\begin{figure}[t] 
  \begin{center}
    \includegraphics[height=2.5cm]{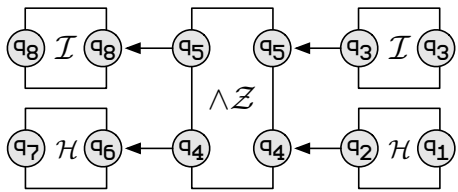}
    \caption{Graphical representation of $\wedge \cX$'s pattern composition.}
    \label{fig:composition}
  \end{center}
\end{figure} 
By simply matching variable names, we can derive
Eq. \eqref{eq:original-cx} automatically from Eq. \eqref{eq:new-cx}.
In fact we can simplify Eq. \eqref{eq:new-cx} further by getting rid
of identity patterns and only expressing non-trivial matching qubit
names, which is specified in our (left-to-right) syntax as follows,
\begin{equation} \label{eq:final-cx}
\{(\cH(q_1,q_2)),\ctR\cZ(q_5,q_4),\cH(q_6,q_7)),
\{(q_2,q_4),(q_4,q_6)\}\}
\text{  .}
\end{equation}
Note that the syntax uses tuples for constraints, and we have reverted
to our left-to-right evaluation order as everywhere in the virtual
model.

In our approach the programmer has to list patterns in order and give
pairs of qubits to specify how patterns are to be composed. Note that
if no pairs are specified this corresponds qubits not being linked up,
or in other words to a parallel composition rather than a sequential
one.  One can then compile pattern composition by instantiating the
involved measurement patterns with fresh variable names and
subsequently matching the correct variable pairs. Now that we have
this generalised notation for pattern composition in place, we can
derive an automatic renaming procedure corresponding to the standard
rules for composing patterns (e.g. linking up the first pattern's
output qubits one by one with the second pattern's input qubits for
sequential composition). This allows us to automate the tedious part
of the qubit renaming process, while at the same time reducing the
chance for human error on larger pattern compositions. However, at the
same time more arbitrary composition structures may still be defined
through the qubit pair syntax provided to the
programmer. Nevertheless, one should view the latter as a more
advanced use of the QVM, while in most cases qubit bindings are
generated as per the standard case.



Once a set of qubit pairs has been specified for a certain pattern
composition -- either automatically or by the user himself -- the
\emph{automated renaming process} works by generating a set of
bindings that map the qubit variable names in every pattern in the
composition to new ones, such that the chosen names are equal if they
appear in the same pair. This is essentially the same process as one
is assumed to execute manually when composing patterns in the formal
model. To ensure that the construction of new bindings occurs in a
well-defined and finite manner, elementary compositions in a composite
structure are processed in topological order. To be precise, an
arbitrary pattern composition expression such as the one in
Eq.~\eqref{eq:final-cx}, is viewed as a graph, with patterns as nodes
and qubit variable pairs as edges. Since we allow only pattern
combinations that form directed acyclic graphs, there is a unique
topological ordering on the list of nodes, and this is the order in
which we evaluate composition bindings. The full set of bindings $B$
is constructed iteratively through the following rules on the pairs of
every elementary pattern composition in topological order.
\begin{equation} \label{eq:b1} 
\inferrule {B(q) = B(q') = \emptyset\;, q_c \text{  fresh}} {
     (q,q') \rightarrow B[q_c/q][q_c/q']}
\end{equation}

\begin{equation}
\label{eq:b2}
\inferrule
{B(q) = q_c}
{(q,q') \rightarrow B[q_c/q']}
\end{equation}

\begin{equation}
\label{eq:b3}
\inferrule
{B(q') = q_c}
{(q,q') \rightarrow B[q_c/q]}
\end{equation}
In other words, when both names in the pair do not appear in the
binding list, rule~\eqref{eq:b1} will trigger. A fresh qubit variable
name $q_c$ is chosen and added as binding for both variable names in
the pair. Rules~\eqref{eq:b2}~\&~\eqref{eq:b3} ensure that if a
binding already exists for one of the variable names in the pair, the
other will use the same binding. The topological sort will ensure that
at all times only one of the three rules will execute.

Once the binding set $B$ is constructed qubit variables in each of the
composing patterns are substituted with their bound value, $P'_i =
\{B(q) | \forall q \in P_i\}$. After this renaming process all
patterns in the composition can finally be merged pairwise by joining
qubit sets in the right way and appending command sequences. Since our
composition mechanism encompasses both definitions in the formal model
our rules for joining qubits sets are also slightly more complicated
(as they are more general). Concretely, we have the following
definition, where patterns are assumed to have passed the renaming
process already.

\begin{definition}
  The \emph{composition} of patterns
  $\cP_{1}=(V_{1},I_{1},O_{1},\cA_{1})$ and $\cP_{2}=(V_{2},I_{2},$
  $O_{2},\cA_{2})$ is defined as the pattern $\cP=(V_1\cup
  V_2,I,O,\cA_{1}\cA_{2})$ where

 \begin{align}
\label{eq:comb2-1}
I &= I_1 \cup (I_2 \backslash O_{1})\\
\label{eq:comb2-2}
O &= (O_1 \backslash I_{2}) \cup O_2\text{  .}
\end{align}
\end{definition}

Indeed, qubit variables from the old input and output sets that were
matched become auxiliary qubits and hence are no longer represented in
the final input and output set.

We stress again that except in the most complex cases, the programmer
does not have to explicitly denote qubit bindings when creating
pattern compositions. Indeed, in most situations these can be created
automatically or one can use implicit notation for qubit bindings as
in the formal model.  For example, we automatically derive regular
sequential composition as specified in Eq.~\eqref{eq:composition} when
both patterns have distinct and unique variable names while the number
of output and input qubits are the same. The automatic renaming
process ensures that the condition $I_2 = O_1$ holds by automatically
creating a composition expression of $P_1$ and $P_2$ while identifying
each qubit in $O_{1}$ with the corresponding one in $I_{2}$ (i.e.  in
sequential order).  Likewise, parallel composition as specified in Eq.~\eqref{eq:tensoring} can be trivially
derived, simply by not specifying any qubit pairs.  Without being
renamed, the qubit names of each pattern in the parallel composition
remain unique. Because of this, Eqs.~\eqref{eq:comb2-1}-\eqref{eq:comb2-2}  essentially
merge the corresponding qubit spaces in the right way.  

With these shortcuts for the original composition operations in hand, we may now return to our example pattern $\ctR \cX$. Indeed, while one way of specifying this composition is given in Eq.~\eqref{eq:final-cx}, by relying on the shortcut for sequential composition we can also express this composition  without qubit names as:
\begin{equation} \label{eq:auto-cx}
\ctR \cX := (\cI \ox \cH) \circ \ctR\cZ \circ  (\cI \ox \cH)
\text{ .}
\end{equation} 
The compiler will instantiate each pattern with fresh variable names, which, since they are unique, will precisely lead to the desired compositions as specified in the expressions above.

\subsection{The distributed quantum virtual machine}\label{sec:dqvm}

Now that our virtual machine for sequential MC computations is in place, we move on to its extension into a distributed version for the DMC language elaborated earlier in this article. Incorporating distribution into our framework comes down to extending each of the abstraction layers.  Our philosophy has been to view network definitions as essentially specialised forms of pattern composition. That is, they are dealt with in a generalisation of the composition layer from Sec.~\ref{sec:compilation}, and compiled towards a set of distributed pattern definitions, i.e. patterns which also allow communication commands. This set of distributed patterns, together with information on how the network is set up in terms of shared resources and agent channels agents,  is then assembled to an executable form and sent to a distributed extension of the execution layer from Sec.~\ref{sec:exec-mc}. We discuss both the execution layer and the composition or \emph{network layer} of the distributed quantum virtual machine (DQVM in short) in Secs.~\ref{sec:dexecution} and \ref{sec:dcomposition} below, first giving  a quick overview here.

In the formal model agents are defined through a pattern with distribution extensions, while the shared resources may be viewed as a
regular non-distributed pattern.  Before discussing the full virtual syntax below, we give the
example of  the leader agent $\bL$ in the entanglement swapping protocol in Eq.~\eqref{esn}. Its data
representation in our QVM is as follows:

\begin{align} \label{eq:esn-leader}
\bL :=  \mtt{(} &\mtt{(?q_0,?q_1,?q_2) \;(?ch_0,?ch_1,?ch_2)}  \notag\\
&\mtt{(} M_{?q_0,?q_1,?q_2}^{GHZ} \:\mtt{(send \; ?ch_0 \; (s \; ?q_0)) \; (send \; ?ch_1 \; (s \; ?q_1)) \; 
(send \; ?ch_2 \; (s \; ?q_2))\mtt{))}} \text{  .}
\end{align}
As before, an agent pattern consists of a qubit sort, specifying which qubits it owns, and a command sequence in which pattern commands as well as communication operations (here \texttt{send} instead of \ttt{!}) can be used. The main difference we see is that agents now also have a \emph{channel sort}, in line with the resource sensitivities of distributed computations and with our naming mechanisms for variables, be it qubits or channels. 

A network definition is essentially a grouping of multiple agent patterns (the single bar $|$) together with a shared resource pattern (the double bar $||$). In our setting we have to augment this with a so-called \emph{network configuration}, a list of qubit and channel variable pairs to organise composition of the network components in the right way. That is, qubit pairs specify how the shared resource pattern needs to compose with the different agents patterns, as in  Sec.~\ref{sec:compilation}, while channel pairs indicate how communication between agents is organised.  Concretely, a network definition is compiled into a list of multiple command sequences that are to be executed concurrently.  The network configuration will enable the automatic renaming process to match the names of various qubit and channel names in these command sequences. We explain this process in more detail in Sec.~\ref{sec:dcomposition} below.


\subsubsection{Execution layer \label{sec:dexecution}} 

The language understood by the execution layer of our distributed quantum virtual machine (DQVM in short) is the lowest-level language in our framework.  It is arrived at after a list of distributed patterns, compiled from a network definition at a higher abstraction layer, is in turn assembled to a list of command sequences devoid of variable names, much as in Sec.~\ref{sec:exec-mc}. Each of these sequences is a data representation of an
agent's command sequence in a distributed network, and the interpreter for DMC needs to run these sequences
concurrently. Note that the shared resource pattern (behind the double bars $||$) is dealt with during a compilation step executed prior to arriving at the execution layer,  so that qubits are initialised in the right way once we are ready for actual execution. While the agent abstraction is not explicitly present in the execution layer, it does provide the required distributed functionality, namely
channel communication operations and concurrent execution of multiple sequences. 

We extend the input language syntax for the QVM (specified in Eq. \eqref{eq:mc-bnf}) to support
DMC computations as follows:

\begin{equation}
  \begin{tabular}{l l l}
\nterm{<network program>}   & ::= &\term{(} \nterm{\{ <agent sequence> \}} \term{)}\\
\nterm{<agent sequence>}    & ::= &\term{(}\nterm{ \{ <agent instruction> \} }\term{)}\\
\nterm{<agent instruction>} & ::= &\nterm{<instruction> |}\\
                                                     &   &\nterm{<channel-send> | <channel-receive>}\\
\nterm{<channel-send>}      & ::= &\term{( send} \nterm{<channel-name> <signal>}   \term{)}\\
\nterm{<channel-receive>}   & ::= &\term{( recv} \nterm{<channel-name> <input-name>} \term{)}\text{  .}\\
 \end{tabular}
\end{equation}
This is again a mirror of the formal syntax, bar the use of the labels \ttt{send} and \ttt{recv} instead of \ttt{!} and \ttt{?} to indicate message sending and reception respectively -- this to avoid confusion with variable names. Also communication commands now carry an extra variable specifying the channel to be used for communication, an improvement over the formal model where the use of channels was somewhat implicit. Finally, we see no syntax for the entanglement resource, which is instead viewed as part of the initialisation procedure carried out in the network layer of our virtual machine. That is, during compilation from the network layer we first run the pattern to construct the shared resources through the QVM.  The quantum state resulting from this execution is then passed on to the initial environment of the execution layer.

The main functionality of the DQVM's interpreter over the QVM's is to do with communication
operations and the scheduling of multiple command sequences.  Practically, this means that the current state of execution retained in the interpreter's environment not only contains information on the quantum and classical states of the network ($Q$ and $\Ga$ respectively) but also on issues to do with channel usage and scheduling. We call the latter the  \emph{network state} $N$ and add it to the interpreter's environment. The network state  $N = \{C,P\}$ holds the \textit{channel map $C$} and the \textit{network program $P$}.  The
latter is a simple collection of each agent's command sequence at each point in the computation, and is used by he interpreter to keep track of each command sequence as it is dynamically scheduled for execution.  Each command sequence is
executed to the point where it is empty or a send or receive operation
blocks.  A next command sequence will then take turn via a round robin
selection and be executed in the same fashion. This round robin system
ensures that sequences with blocking operations become available for execution at later time, successfully executing the send or receive
operation in question if a matching receive or send has been performed
by a previous sequence's turn. This process halts when all command
sequences are empty. Deadlocks or similar errors will cause the
interpreter to get stuck in a loop. We note that the formal semantics of a network is the same for any chosen schedule and for this reason we may choose a schedule at will without having to worry about the network's behaviour deviating from the intended one~\citep{Danos05b}. 

The second component of the network state is the channel map $C$. It simply maps channel names to values (sent along the channel) and
is used to implement the communication primitives in a straightforward way:

\begin{equation}
\inferrule
{C(ch) = \emptyset\\ \Ga(s)=v}
{\Ga , C ,  (\ttt{(send}\; ch \; s \ttt{)} \; \cE) \\
  \rightarrow \\ 
\Ga , C[ v/ch] , ( \cE ) }
\end{equation}
\begin{equation}
\inferrule
{C(ch)=v}
{\Gamma , C, 
(\ttt{(recv} \;ch \; name \ttt{)} \; \cE) \\
  \rightarrow \\ 
\Gamma[v/name], C[\emptyset/ch], ( \cE ) }\text{  .}
\end{equation}


Here $ch$ is a channel name, $\Gamma(s)$ denotes the value of the
signal $s$ with respect to the given outcome map and \cE is the rest of the
execution sequence. In other words, the send operation associates the value to be sent to the appropriate channel name in the channel map $C$, while the
corresponding receive operation removes the value from $C$ and stores it in the classical state. In case where a channel already contains a value
during a send operation along a particular channel, the send operation blocks, and likewise for a receive operation on an empty channel:
\begin{equation}
\inferrule
{C(ch)=v}
{C , 
(\ttt{(send} \;ch \; v' \ttt{)} \; \cE) \\
  \rightarrow \\ 
C,   (\ttt{(send} \;ch \; v'
\ttt{)} \; \cE)}
\end{equation}

\begin{equation}
  \inferrule {C(ch) = \emptyset }
  {C , (\ttt{(recv} \;ch \; name \ttt{)} \; \cE) \\
  \rightarrow \\ 
C , (\ttt{(recv} \;ch \; name \ttt{)} \; \cE)}\text{  .}
\end{equation}
Note that these rules constitute a concretisation of the semantical rule for classical communication in the formal model (see Eq.~\ref{rendvcl}). Here we have chosen a concrete semantics using a form of \emph{blocking}, in the sense that execution of a particular command sequence may halt until certain requirements are met, embodied in the equations above.

Next to the network state the interpreter's environment also keeps track of the quantum and the classical state of the network. For practical reasons we reuse the original QVM's computation state as described in Sec.~\ref{sec:qvm}, i.e. we have one global quantum state $Q$ and one global classical state
$\Gamma$ for the entire network. This simplifies the structure of the DQVM, allowing it to focus on the network-specific features while the QVM is used
for the execution of regular MC operations.  Indeed, the quantum state $Q$ is passed through unchanged to the QVM which executes regular MC (i.e. quantum) operations.  The execution schedule described above imposes a synchronised access to this global state $(Q,\Gamma)$, while our naming procedure ensures that names are unique so that there are no clashes between classical variable names of different agents. Note that a more sophisticated implementation would adhere to the DMC's formal semantics where each agent has its own local state.  For local quantum states this is to some extent captured by our use of tangles for disentangled quantum states (cfr. Eq.~\eqref{eq:qstate}). For classical states this would require a separate environment for each agent such that a received value is stored in the local classical state of the receiving agent.

\subsubsection{Network layer} \label{sec:dcomposition}

The network layer is the place where the programmer defines distributed programs, such as for example the network for entanglement sharing specified formally in Eq.~\eqref{esn}. A network consists of a list of agent definitions, a shared resource pattern, and a network configuration which specifies how resources are distributed and connected. Concretely we need to specify a list of patterns, namely the extended
patterns of all involved agents and a single regular pattern for the shared resources.  For example, to define the ES network from Eq.~\eqref{esn} we need the leader agent pattern given in Eq.~\ref{eq:esn-leader} together with the agent pattern $\bA$  and the resource pattern $\cR$:

\begin{align} \label{eq:esn-agents}
\bA& :=  \mtt{( (?q) \;(?ch) \;  ((recv \; ?ch \; (s \; ?v))\;(X \; ?q \;(s \; ?v))))}\\
 \cR &:=\mtt{(V \quad I \quad O \quad((E \; ?a \; ?b) \; (E \; ?c \; ?d)\;  (E \; ?e \; ?f)))} \text{ ,}
\end{align}
where \ttt{V = I = O = (?a ?b ?c ?d ?e ?f)}. The full network is specified by the expression
$ES := ( \cR , \ens{\bL, \bA,\bA,\bA}, NC) \text{ ,}$
where $NC$ is the network configuration. As in Sec.~\ref{sec:compilation}, each of these
patterns is instantiated with distinct qubit and channel variable names.  The network configuration specified drives the composition of patterns in a network, which steers an automatic renaming process much as in Sec.~\ref{sec:compilation}.  This network configuration contains two things. First, a list
of qubit variable pairs linking output qubit variables of the shared resource pattern to input qubit variables of some agent pattern. Second, besides qubit variables there are now also channel variables to be matched between agent patterns, due to the communication extension to their command sequence. However, matching of channel names can be performed by the same renaming process defined earlier. 

After the automated renaming process has finished, agent patterns are assembled and collected in a list, forming the multiple command
sequences list that is used as input to the DQVM's interpreter. The pattern for shared resources does not appear in the command sequences list for the DQVM.  Instead, it is assembled and executed by the QVM as an the initialisation step, and passed on as initial quantum state $Q$ in the environment to the DQVM when it starts executing the agent programs of the network.

\subsection{A graphical user interface\label{sec:graphics}}

In the above we gave an overview on the design and architecture of our virtual machine for the DMC language. While the subject of this article has to do with the inner structure of our framework, the goal of the latter is nevertheless to provide a user-friendly programming environment. For this reason we have developed a graphical user interface (GUI) on top of the virtual machine to facilitate experimentation. This GUI tool currently supports only regular (non-distributed) patterns, allowing definition of patterns from scratch as well as in terms of compositions of existing patterns. Figure~\ref{fig:gui-tool} shows the GUI being used to define a pattern, the $W_3$ entanglement pattern, as a composition of known patterns; at the bottom we find the compiled version of that same pattern.  As seen previously in this section our framework allows for a general, more explicit form of composition, such that explicit set of qubit name pairs subsumes the implicit method of matching qubits by manual rewriting. While expressing these pairs in writing is still a feat for large programs, a graphical notation similar to Figure~\ref{fig:composition} is much more expressive. Our GUI uses a similar notation where black boxes denote in- and outputs and the user may connect these in arbitrary ways to concretise the desired connections. This technique allows a quantum programmer to express complex quantum algorithms more easily, certainly for composition structures with non-trivial connections between component patterns.  

\begin{figure}[ht] \label{fig:gui-tool}
\centering
\includegraphics[height=10cm]{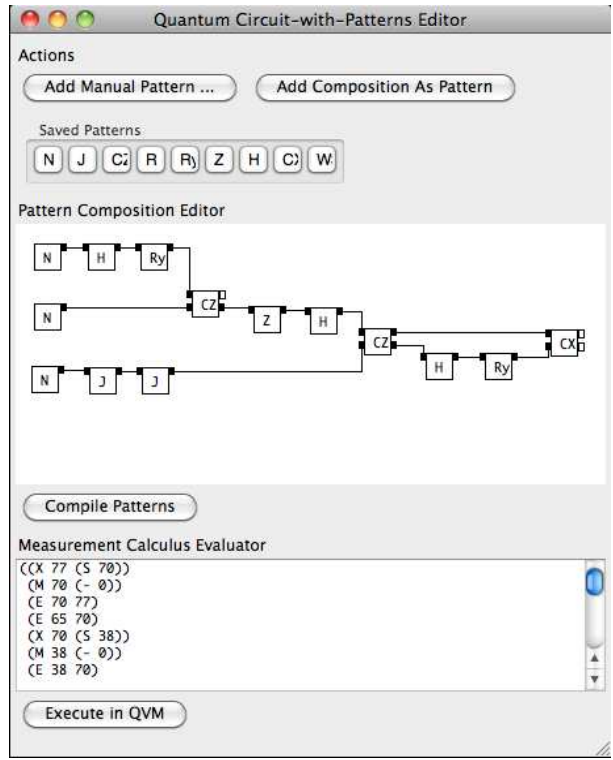}
\caption{Graphical User Interface of the QVM's design tool, showing
  the composition and compilation of the 3-qubit $W$ entanglement
  state preparation pattern.}
\end{figure}

 \section{ Conclusion \label{sec:conclusion}}

In the above we describe an assembly language for distributed meas\-ure\-ment\--based quantum (or DMC) computations in all its aspects. While the first half of this article deals with the formal model, the second half elaborates on a virtual framework developed in close relationship with the formal model, i.e. a programming environment for the DMC language. 

DMC programs satisfy several formal properties crucial to the practical usability of the language, such as compositionality and context-freeness. We showed how to put these properties to use by formally implementing a composite program to control operations in a distributed setting, and demonstrated that the semantics  does not change under the various composition operations. DMC was developed with expressiveness in mind, a crucial property when establishing the first layer in a distributed quantum programming paradigm. Indeed only through experimentation combined with abstraction can one hope to move towards a higher level in the language hierarchy. Our first experiments on paper already prove that DMC is indeed capable of  expressing more complicated programs and that the formal features are necessary and sufficient in determining their functionality. This is very different in flavour from earlier formal frameworks in this area~\citep{Gay04,Jorrand05}, which are much more concerned with issues such as verification, and focus on providing tools such as type systems to facilitate investigations of this nature. 

While the first half or this article shows that it is possible in principle to write DMC programs within the formal model, the limits of its usability are acutely felt even for the relatively small example of distributed controlled operations. Indeed, as an experimentation platform our formal framework is only really useful in terms of a quantum virtual machine (QVM), a programming environment for the DMC language automating execution and composition of DMC programs. In the second half of this work we discuss a first implementation of such a virtual machine, a proof-of-concept execution environment effectively virtualising the formal DMC model. The benefits of having a QVM are many: as a first layer of a tiered quantum computation architecture~\citep{Svore06}, a platform for automated verification and model-checking, and maybe most importantly, a basis from which to develop higher-order distributed quantum computation languages through experimentation and abstraction.  Our QVM is built up in terms of  several abstraction layers, most importantly a platform-independent execution layer to deal with the low-level semantics of basic patterns, and a composition layer to create and compose larger programs. The composition layer comes with an associated compiler that translates any compositional structure to one single pattern definition. Any pattern definition, be it specified directly by the user or produced by the compiler, is assembled into an expression that the execution layer understands, essentially a command sequence where all variable names have been replaced by concrete references in the intended way. While the execution layer is defined independently of any actual implementation, we used a Lisp-based simulation environment in our experiments to evaluate execution-layer expressions. The QVM is extended with distributed structures at all layers into a distributed quantum virtual machine (DQVM) which allows specification and execution of arbitrary distributed networks. Finally, a  graphical user interface (GUI) is added to facilitate usability of the framework.  We note that due to the fact that some aspects of the formal model necessitate further concretisation in a virtual model, there are some differences in syntax and semantics between the two. For example the implicit naming conventions for variables, channels, and in the composition of programs, required a concrete design in the virtual setting. 

The virtual machine developed in this article, while covering almost all aspects of the formal model, is a first implementation and as such there are several avenues for improvement. We list these here moving from higher-abstraction layers to lower ones. First, the distributed layer requires several extensions to be fully compatible with the formal model, most importantly by giving agents a more prominent role and allowing network composition. In order to develop the GUI into a fully-featured development tool, it also needs to be lifted to a distributed setting: adding agents and networks to the graphical notation is top priority. Ultimately the goal is to make the graphical notation into a self-contained language, allowing the programmer to specify arbitrary programs without needing to create ex-nihilo patterns, which involves writing command sequence code by hand. Second, much work can be done at the level of optimisation of the execution layer, which now relies on  conventional rather than optimal data structures for its implementation. We are in the process of developing tailor-built data structures,  so enhancing the performance of our framework by relying on domain-specific measurement calculus optimisations. For example, we are currently looking into the stabiliser calculus as a means to efficiently execute MC operations where possible, switching to a different representation when operations unsupported by the stabiliser calculus are performed. Optimisations at the level of command sequences are also possible, such as detecting and directly initialising cluster states rather than constructing them incrementally by executing the required entanglements. On the other hand, classically simulating large entangled states means dealing with an exponential blow-up of computational time and space. For this reason the so-called EMC form~\citep{Danos04b} (which puts entanglements first) is not always the preferred one in a simulation setting, since we need to  minimise the size of entangled states during the length of the computation. Finding the right balance between direct cluster state generation and exploiting classical resources at their fullest is an exercise which is currently underway. Finally, we are heavily looking into parallel computing techniques to improve the simulation of quantum operations, at the moment carried out by straightforward linear algebraic techniques.  Concretely, we are investigating the compilation of command sequences into a dataflow network~\citep{Gordon:2006p38}. In such a network quantum states are represented by a long stream of amplitudes, which has the double benefit of exposing the inherent parallelism while at the same time relaxing the need to fit entire vectors inside the same computer memory. This line of research has already lead to a first implementation of a parallelised simulation environment in ~\citep{Verhaegen09}.


\section{Acknowledgements}
This work is based on earlier research with V. Danos, E. Kashefi and P. Panangaden.  The first author is supported by the Flemish Fund for Scientific Research (FWO) and the second by the CRYPTASC project, funded by the Institute for the encouragement of Scientific Research and Innovation of Brussels (IRSIB/IWOIB).

\section{Appendix}

Here we give the full operational semantics of the distributed measurement calculus. We are concerned here with the small-step rules, indicating how atomic expressions in the language are evaluated. The big-step operational semantics of a program is found by grouping the pertaining small-step rules into one transition for the whole program. We use the standard notation of sequents and rules. A sequent $\Ga \ent E\dar $ is read as ``given environment $\Ga$, the expression $E$ evaluates to the value $v$''.  In case that the environment itself changes during evaluation of an expression $E$ we write $\Ga, E \lrar \Ga'$. We write $\Ga(x)$ for the value of $x$ in $\Ga$ and  $\Ga[v/x]$ for the environment $\Ga$ with the added binding of $x$ to $v$. Sequents can be combined into rules $\frac{S}{S'}$ which are just a different notation for $S \Rar S'$.

There are four groups of rules, dealing with classical values (signals and angles), measurement patterns and distributed measurement patterns respectively. Each group builds on top of the previous one. The first group of rules is to do with the evaluation of signals.

\GA{
\frac{}{\Ga \ent 0 \dar 0} \quad \text{and} \quad \frac{}{\Ga \ent 1 \dar 1} \label{signal1}\\
\frac{}{\Ga \ent s_{i} \dar \Ga(i)} \label{signal2}\\
\frac{}{\Ga[v/s_{i}] \ent s_{i} \dar v }  \quad \text{and} \quad \frac{}{\Ga[v/s_{i}] \ent s_{j} \dar \Ga(j)} \quad \text{ if } i\neq j  \label{signal3}\\
\frac{\Ga \ent s \dar v \quad   \Ga \ent t \dar u}{\Ga \ent s+t \dar v\oa u}\label{signal4}
}
Here $\Ga$ is the outcome map or classical state, and $\oa$ denotes addition in $\mbb{Z} _{2}$.  Angles, which can have signal dependencies percolated through via dependent measurements, are also purely classical. Values of signals are looked up in $\Ga$ via the ruleset for signals in order to determine the actual value of a measurement angle. This procedure is summarised in the following rules.

\GA{
\frac{}{\Ga \ent \al \dar \al}\\
\frac{\Ga \ent s \dar v \quad \Ga \ent t \dar u}{\Ga \ent \a st \dar (-1)^{v}.\al+u.\pi}\\
\frac{}{\Ga \ent \as s \dar \a s 0}  \quad \text{and} \quad  \frac{}{\Ga \ent \at t \dar \a 0 t}
}

Having defined how signals and angles are evaluated, we can now move on to the operational semantics of the basic commands. These commands operate on an environment consisting of a quantum state $\rho$ as well as a classical state $\Ga$. We have presented these rules here for density matrices; in pure state derivations we often use state transitions for brevity. Specifically, for a pure state we have $\rho=\proj{\psi}$, which is mapped to $L\proj{\psi}L^{\dag}$, with $L$ any of the entanglement, Pauli or projection operators below. A pure state transition can then be alternatively specified as mapping $\gket$ to $L\gket$.

\GA{
\frac{}{\rho, \Ga , \et i j \lrar \ctR Z_{ij}\rho  \ctR Z_{ij} ,\Ga}\\
\frac{\Ga \ent s \dar v} {\rho, \Ga , \cx is  \lrar   \cx i v \rho \cx i v ,\Ga   }\\
\frac{\Ga \ent s \dar v} {\rho, \Ga , \cz is  \lrar   \cz i v \rho \cz i v, \Ga }\\
\frac{\Ga \ent \a st \dar \ba}
 {\rho, \Ga,\MS{\al}ist  \lrar_{\la_{0}} \pp\ba \rho\: \ket{+_{\ba}}_{i}, \Ga[0/i] } \mbox{\quad, } \la_{0}=\frac{\mrm{tr}( \proj{+_{\ba}}_{i}\:\rho)}{\mrm{tr} \rho} \label{meas1}\\
\frac{\Ga \ent \a st \dar \ba}
 {\rho, \Ga,\MS{\al}ist  \lrar_{\la_{1}}\mp\ba \rho\:\ket{-_{\ba}}_{i}, \Ga[1/i] }   \mbox{\quad, } \la_{1}=\frac{\mrm{tr}(\proj{-_{\ba}}_{i}\:\rho)}{\mrm{tr} \rho}\label{meas2}\\
 \frac{\rho, \Ga, C_{1} \lrar_{\la} \rho', \Ga'}
 {\rho, \Ga,C_{2}C_{1}  \lrar_{\la}\rho', \Ga',C_{2} }
 }
The first three commands are purely quantum and straightforward. The measurement command is the only command that affects the quantum state as well as the classical state. First, the measurement angle has to be evaluated, which in turn requires evaluating the $X$- and $Z$-signals by the previous sets of rules. Measurement commands are also the only nondeterministic commands, as the measured qubit is projected onto either $ \ket{+_{\al}}$ or $ \ket{-_{\al}}$ with transition probabilities as stated. Usually, the convention is to renormalise the state after measurement, but we do not adhere to it here, as in this way the probability of reaching a given state can be read off its norm, and moreover the overall treatment is simpler. The last rule is for a composition of commands. 

Finally, we need a set of rules to deal with distributed extensions to patterns. Essentially these transitions describe how agents $\bA(i,o): Q.\cE$ and networks $\cN=|_{i}\bA_{i}(\bi_{i},\bo_{i}):Q_{i}.\cE_{i}\given \sig $ evolve over different time steps.  We adopt a shorthand notation for agents, leaving out classical inputs and output, which do not change with small-step reductions.

\EQ{\SP{ 
\bfa_{i}&= \bA_{i}:Q_{i}.\cE_{i}\\
\bfa_{i}.E & = \bA_{i}:Q_{i}.[\cE_{i}.E] \\
\bfa^{-q}&=\bA:Q\bs\ens{q}.\cE\\ 
 \bfa^{+q}&= \bA:Q\uplus\ens{q}.\cE[q/x] \text{  ,}
}}
 where $E$ is some event, and $\cE_{i}$ and $\cE'_{i}$ are event sequences.  A \emph{configuration} is given by the  system state $\sig$ together with a set of agent programs, and their states, specifically
\EQ{ 
\sig, |_{i}\Ga_{i},\bfa_{i}=\sig, \Ga_{1},\bfa_{1}\nr \Ga_{2},\bfa_{2}\nr\dots \nr  \Ga_{m},\bfa_{m}\text{  .}
\label{configs}
}
The small-step rules for configuration transitions, denoted $\Lrar$, are specified below; we provide some explanations afterwards. When the system state is not changed in an evaluation step, we stress this by preceding a rule by $\sig \ent$. 

\GA{
 \frac{\sig, \cP(V,I,O, \cA),\Ga \lrar_{\la} \sig',\Ga'} {\sig,\Ga,\bA:I\uplus R.[\cE.\cP] \Lrar_{\la} \sig', \Ga',\bA:O\uplus R.\cE} \label{localop}\\
\notag\\
\frac{\Ga_{2}(y)=v}{\sig \ent ( \Ga_{1},\bfa_{1}.\cc?x \nr \Ga_{2}, \bfa_{2}.\cc!y \Lrar \Ga_{1}[x \mapsto v],\bfa_{1} \nr \Ga_{2},\bfa_{2})} \label{apprendvcl}\\
\notag\\
 \frac{}{\sig \ent ( \Ga_{1},\bfa_{1}.\qc?x \nr \Ga_{2},\bfa_{2}.\qc!q \Lrar \Ga_{1},\bfa_{1}^{+q}\nr \Ga_{2},\bfa_{2}^{-q})} \label{apprendvq} \\
\notag\\
 \frac{L \Lrar_{\la}M}{L \nr N \Lrar_{\la} M \nr N} \label{metarule}
}

Implicit in these rules is a sequential composition rule, which ensures that all events in an agent's event sequence are executed one after the other.  The first rule is for local operations; we have written the full pattern instead of only its command sequence here to make pattern input and output explicit. Because a pattern's big-step semantics is given by a probabilistic transition system described by $\lrar$, we pick up a probability $\la$ here.  Furthermore, an agent changes its sort depending on pattern's output $O$. The next rule is for classical rendezvous and is straightforward. For quantum rendezvous, we need to substitute $q$ for $x$ in the event sequence of the receiving agent, and furthermore adapt qubit sorts. The last rule is a metarule, which is required to express that any of the other rules may fire in the context of a larger system. $L$ and $R$ stand for any of the possible left-, respectively right-hand sides of any of the previous rules, while $L'$ is an arbitrary configuration. Note that we might need to rearrange terms in the parallel composition of agents in order to be able to apply the context rule.  This can always be done since the order of agents in a configuration is arbitrary. In derivations of network execution, we often do not explicitly write reductions as specified by \eqref{metarule}, but rather specify in which order the other rules fire for the network at hand. It is precisely in this last rule that introduces nondeterminism at the network level, that is, several agent transitions may be possible within the context of a network at the same time.

\setcitestyle{authoryear,round,comma}
\bibliography{references}
\bibliographystyle{natbib}

\end{document}